\documentclass[aps,prl,superscriptaddress,twocolumn,showpacs,preprintnumbers]{revtex4}
\usepackage[tbtags]{amsmath}
\usepackage{amsfonts}
\usepackage{mathrsfs}
\usepackage{amssymb}
\usepackage{graphicx}
\usepackage{epsfig,graphicx,times}
\usepackage{subfigure}
\usepackage{color}
\usepackage{amssymb}
\usepackage{amsmath}
\usepackage{graphicx}
\usepackage{dcolumn}
\usepackage{bm}
\usepackage{float}
\usepackage[abs]{overpic}

\begin{document}

\title{Probing molecular chirality by coherent optical absorption spectra}
\author{W. Z. Jia}
\affiliation{Quantum Optoelectronics Laboratory, School of Physics and Technology,
Southwest Jiaotong University, Chengdu 610031, China}
\author{L. F. Wei}
\affiliation{Quantum Optoelectronics Laboratory, School of Physics and Technology,
Southwest Jiaotong University, Chengdu 610031, China}
\affiliation{State Key Laboratory of Optoelectronic Materials and Technologies, School of
Physics and Engineering, Sun Yat-Sen University, Guangzhou 510275, China}

\begin{abstract}
We propose an approach to sensitively probe the chirality of
molecules by measuring their coherent optical absorption spectra. It
is shown that quantum dynamics of the cyclic three-level chiral
molecules driven by appropriately-designed external fields is
total-phase dependent. This will result in chirality-dependent
absorption spectra for the probe field. As a consequence, these
absorption spectra can be utilized to identify molecular chirality
and determinate enantiomer excess. The feasibility of the proposal
with chiral molecules confined in hollow-core photonic crystal fiber
(HC-PCF) is also discussed.
\end{abstract}

\maketitle
\email{jiawenz1979@126.com}
\email{lfwei@swjtu.edu.cn}


\bigskip

\textit{Introduction.---}The coexistence of left- and right-handed
chiral molecules (called \textquotedblleft
enantiomers\textquotedblright ) originates from the fundamental
broken symmetries in nature~\cite{chirality}. The physiological
effects of enantiomers of biologically active compounds may differ
significantly~\cite{differLR}. In general, only one enantiomeric
form has the potential to be biologically beneficial, while the
other one could be harmful or fatal. Thus, probing molecular
chirality is fundamentally important tasks in organic chemistry,
pharmacology, biochemistry, etc..

Since Pasteur's pioneering experiments on optical
activity~\cite{Pasteur}, asymmetries in the interaction of polarized
light with chiral molecules has provided a powerful physical probe
of molecular chirality. The most common techniques providing
chirality-specific spectroscopic information are, e.g., circular
dichroism (CD), and Raman optical activity (ROA),
etc.~\cite{conventional}.
Note that all these techniques are based on certain higher-order
interaction (such as magnetic-dipole and electric-quadrupole
couplings) between the molecular systems and the probing lights.
Consequently, the sensitivity of these conventional techniques is
usually not sufficient to detect chirality of a small amount of
sample. While, recent studies~\cite{SHG1,SHG2,SFG1,SFG2} show that,
purely electric-dipolar nonlinear optical process, such as
second-harmonic generation (SHG) and sum-frequency generation (SFG),
can act as desirable sensitive probes of chirality.

Alternatively, in this letter we propose an optical method to
achieve the desirable chirality probe. Our proposal is based on
manipulating quantum coherence in chiral molecules, which can be
modeled as cyclic-transition ($\Delta$-type) quantum systems
~\cite{Shapiroprl87,Shapiroprl90,LYX}. Under strong driving the
coexistence of electromagnetic induced transparency (EIT) effects
~\cite{EIT} and two-photon process in a $\Delta$-type chiral
molecule will result in chirality-dependent absorption spectra for
the applied probe field. In fact, based on this kind of transition
structure some optical methods for enantioseparation (rather than
chirality probes) have already been
suggested~\cite{Shapiroprl87,Shapiroprl90,LYprl,dynamicalcontrol}.
Since our protocol is based on the electric-dipole interactions
between the molecules and the coherent optical fields, it may
provide a more sensitive chirality probe compared with the
conventional ones by measuring the optical
activity~\cite{conventional}.

\begin{figure}[b]
\includegraphics[width=0.4\textwidth]{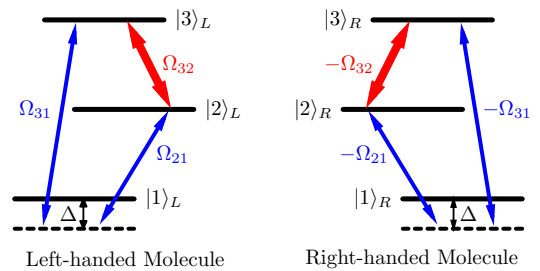}
\caption{(Color online) Three-level chiral molecules with cyclic
transition structures. The enantiomers are driven by three optical
fields. Levels $\left\vert 2\right\rangle _{\mathrm{L}}$
($\left\vert 2\right\rangle _{ \mathrm{R}}$) and $\left\vert
3\right\rangle _{\mathrm{L}}$ ($\left\vert 3\right\rangle
_{\mathrm{R}}$) are resonantly coupled by a strong control field.
The other two transition channels are coupled by two probes with the
same detuning $\Delta $.} \label{EL}
\end{figure}

\textit{Chirality probe based on manipulating quantum coherence.---}
It is well-known that a pair of chiral molecules (e.g., the
$D_{2}S_{2}$ enantiomers) can be modeled as a system with
mirror-symmetric double-well potential~\cite{Shapiroprl90,Hund}. Due
to the broken-parity symmetries, the lowest three localized chiral
eigenstates can form a $\Delta$-type cyclic-transition
structure~\cite{Shapiroprl87,Shapiroprl90}. We assume that
$\left\vert i\right\rangle $ ($i=1,2,3$) are the selected levels of
a chiral molecule with according eigenfrequenies $\omega _{i}$.
Specifically, $\left\vert i\right\rangle$ may be left-handed states
$\left\vert i\right\rangle _{\mathrm{L}}$ or right-handed states
$\left\vert i\right\rangle _{\mathrm{R}}$, as shown in
Fig.~\ref{EL}. Three coherent driving fields $E_{ij}\left(
z,t\right) =\frac{1}{2}\mathscr{E}_{ij}\left( z,t\right) e^{-i\left[
\nu _{ij}t-k_{ij}z+\phi _{ij}\left( z,t\right) \right]
}+\mathrm{c.c.}$ ($i>j$) with slowly varying amplitude
$\mathscr{E}_{ij}$, wave vectors $k_{ij}$, frequencies $\nu _{ij}$
and phase factor $\phi _{ij}$, are applied to couple all the
permissible transition-channels $\left\vert i\right\rangle
\leftrightarrow \left\vert j\right\rangle $.

The Hamiltonian describing the present three-level cyclic-transition
system can be written as
\begin{eqnarray}
\hat{H}=\hbar \sum_{i=1}^{3}\omega _{i}\left\vert i\right\rangle
\left\langle i\right\vert-\frac{\hbar }{2}\sum _{i>j=1}^{3}\left(
\Omega _{ij}e^{-i\nu _{ij}t+ik_{ij}z}\left\vert i\right\rangle
\left\langle j\right\vert +\mathrm{H.c.}\right).
\end{eqnarray}
Here, the Rabi frequencies are defined as $\Omega _{ij}=\left( \mu
_{ij}\mathscr{E}_{ij}e^{-i\theta _{ij}}\right) /\hbar$, with $\mu
_{ij}$ being the amplitude of the dipole matrix elements and $\theta
_{ij}=\phi _{ij}+\chi _{ij}$. $\phi _{ij}$ and $\chi_{ij}$ are the
phases of the electric field components and the dipole matrix
elements, respectively.
The master equation describing the dynamical evolution of the system
can be written as
\begin{equation}
\frac{d\hat{\rho}}{dt}=-\frac{i}{\hbar }
[\hat{H},\hat{\rho}]+\mathcal{L}[\hat{\rho}].
\end{equation}

We now assume that the strong control field $\Omega_{32}$ is
resonantly applied and the two weak probe fields, $\Omega_{31}$ and
$\Omega_{21}$, are applied with the same detunings, namely, $\Delta
_{32}=0$, $\Delta _{31}=\Delta _{21}=\Delta$. Then, the condition
$\nu _{31}=\nu _{21}+\nu _{32}$ and consequently
$k_{31}=k_{21}+k_{32}$ are satisfied (Here, three fields are assumed
to be propagating in the same direction). Furthermore, we redefine
the density matrix elements as $\sigma _{ii}=\rho _{ii}$, $\sigma
_{ij}=\rho _{ij}e^{i\nu _{ij}t-ik_{ij}z+i\theta _{ij}}$, ($i>j$).
Finally, under the rotating wave approximation the redefined density
matrix elements obey the following equations of motion:
\begin{subequations}
\begin{eqnarray}
\dot{\sigma}_{11} &=&\Gamma _{31}\sigma _{33}+\Gamma _{21}\sigma _{22}+\frac{%
1}{2}(-i\left\vert \Omega _{31}\right\vert \sigma _{13}  \notag \\
&&-i\left\vert \Omega _{21}\right\vert \sigma _{12}+\mathrm{H.c.}),
\label{Mastereq1}
\\
\dot{\sigma}_{22} &=&-\Gamma _{21}\sigma _{22}+\Gamma _{32}\sigma _{33}+%
\frac{1}{2}(i\left\vert \Omega _{21}\right\vert \sigma _{12}  \notag \\
&&-i\left\vert \Omega _{32}\right\vert \sigma _{23}+\mathrm{H.c.}),
\label{Mastereq2}
\\
\dot{\sigma}_{21} &=&-\lambda _{21}\sigma
_{21}+\frac{1}{2}[i\left\vert \Omega _{32}\right\vert \sigma
_{31}e^{i\Theta }-i\left\vert \Omega _{31}\right\vert \sigma
_{23}e^{i\Theta } \notag \\
&&-i\left\vert \Omega _{21}\right\vert \left( \sigma _{22}-\sigma
_{11}\right) ], \label{Mastereq3}
\\
\dot{\sigma}_{31} &=&-\lambda _{31}\sigma
_{31}+\frac{1}{2}[i\left\vert \Omega _{32}\right\vert \sigma
_{21}e^{-i\Theta }-i\left\vert \Omega _{21}\right\vert \sigma
_{32}e^{-i\Theta }  \notag \\
&&-i\left\vert \Omega _{31}\right\vert \left( \sigma _{33}-\sigma
_{11}\right) ], \label{Mastereq4}
\\
\dot{\sigma}_{32} &=&-\gamma _{23}\sigma
_{32}+\frac{1}{2}[i\left\vert \Omega _{31}\right\vert \sigma
_{12}e^{i\Theta }-i\left\vert \Omega _{21}\right\vert \sigma
_{31}e^{i\Theta }  \notag \\
&&-i\left\vert \Omega _{32}\right\vert \left( \sigma _{33}-\sigma
_{22}\right) ], \label{Mastereq5}
\end{eqnarray}
\end{subequations}
where $\Gamma_{ij}$ ($i>j$) are the relaxation rates between the
levels $\left\vert i\right\rangle$ and $\left\vert j\right\rangle$;
$\lambda_{21}=\lambda_{12}^{*}=\gamma _{12}-i\Delta $, $\lambda
_{31}=\lambda_{13}^{*}=\gamma _{13}-i\Delta $;
$\gamma_{ij}=\gamma_{ji}$ are the damping rates of the off-diagonal
terms, and $\Theta =\theta _{32}+\theta _{21}-\theta _{31}$ are the
relative phases.

The above equations implie that the steady-state density matrix
elements $\sigma _{ij}^{\left( s\right) }$ should be phase
dependent~\cite{LHB,Jia}. Additionally, all of the Rabi frequencies
for the two enantiomers could differ by a sign, resulting in a $\pi$
difference of total phase factor~\cite{Shapiroprl87}. Consequently,
the steady-state density matrix elements of the left- and
right-handed molecules can be presented as $\sigma_{ij}^{\left(
+\right) }=\sigma _{ij}^{\left( s\right) }\left( \Theta \right) $
and $\sigma _{ij}^{\left( -\right) }=\sigma _{ij}^{\left( s\right)
}\left( \Theta +\pi \right) $, respectively.

\begin{figure}[t]
\includegraphics[width=0.3\textwidth]{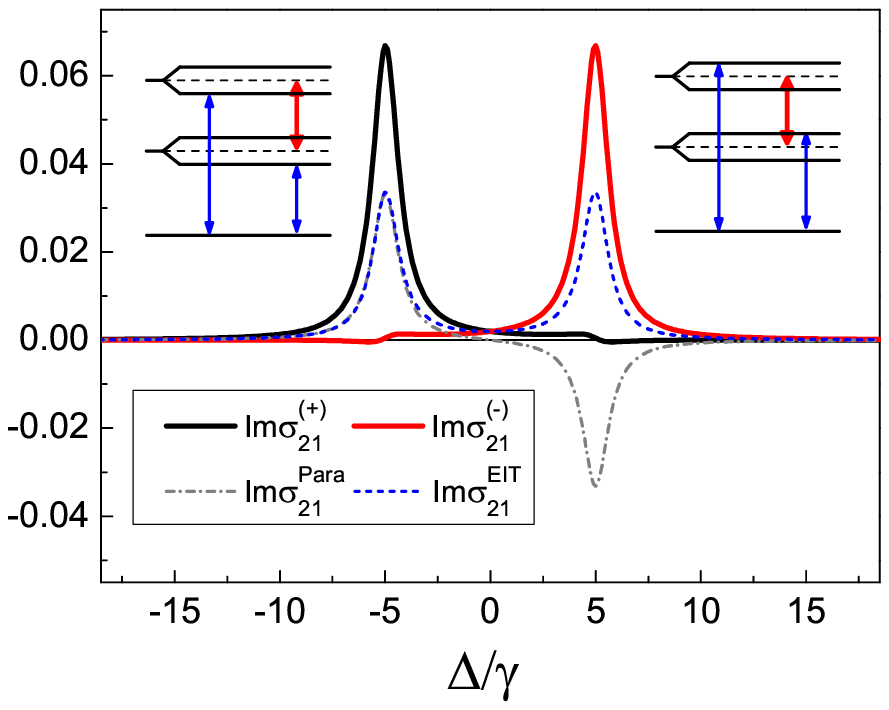}
\caption{(Color online) Parameters $\mathrm{Im}\protect\sigma
_{21}^{\left( \pm \right) }$, $\mathrm{Im}\protect\sigma
_{21}^{\mathrm{EIT}}$, $\mathrm{Im}\protect\sigma
_{21}^{\mathrm{Para}}$ versus the dutuning $\Delta$, showing that
under the condition~\eqref{Condition}, the chirality-dependent
quantity $\mathrm{Im}\protect\sigma _{21}^{\left( +\right) }$
($\mathrm{Im}\protect\sigma_{21}^{\left( -\right) }$) is the
symmetric (antisymmetric) superposition of
$\mathrm{Im}\protect\sigma _{21}^{\mathrm{EIT}}$ and
$\mathrm{Im}\protect\sigma _{21}^{\mathrm{Para}}$. Here, we set
$\Theta =0$, $ \left\vert \Omega _{21}\right\vert =\left\vert \Omega
_{31}\right\vert =\protect\gamma /10$, $\left\vert \Omega
_{32}\right\vert =10\protect\gamma $, $\Gamma _{ij}=\protect\gamma
$, $\protect\gamma _{12}=\Gamma _{21}/2$, $\protect\gamma
_{13}=\left( \Gamma _{31}+\Gamma _{32}\right) /2$, and
$\protect\gamma _{23}=\left( \Gamma _{21}+\Gamma _{31}+\Gamma
_{32}\right) /2$. Insets: schematics of the quantum interference
between the single- and two-photon transition channels at probe
detunings $\Delta =\pm \left\vert \Omega _{32}\right\vert /2$.}
\label{interference}
\end{figure}

This phase sensitive optical response can be utilized to probe
molecular chirality. In our protocol, either of the two weak probes
can be served as the chirality probe, e.g., we can choose the one
with Rabi frequency $\Omega _{21}$ for specific analysis. For this
purpose, we firstly give the analytical expressions of $\sigma
_{21}^{\left( \pm \right) }$ (which represent the induced coherence
between levels $\left\vert 1\right\rangle$ and $\left\vert
2\right\rangle$) in the first order of $\Omega _{21}$ and
$\Omega_{31}$:
\begin{equation}
\sigma _{21}^{\left( \pm \right) }=\frac{i \lambda _{31}\left\vert
\Omega _{21}\right\vert}{2Z}\pm\frac{-\left\vert \Omega
_{31}\right\vert \left\vert \Omega _{32}\right\vert e^{i\Theta
}}{4Z}
\label{sigma21}
\end{equation}
with $Z=\frac{1}{4}\left\vert \Omega _{32}\right\vert ^{2}+\lambda
_{21}\lambda _{31}$. Above, the first term (denoted by
$\sigma_{21}^{\mathrm{EIT}}$) and the second one (denoted by $\sigma
_{21}^{\mathrm{Para}}$) are related to the (Ladder-type) EIT effect
and the two photon parametric process, respectively.
Obviously, due to the intrinsic $\pi$-difference between the two
enantiomers, the resultant phase-dependent interference terms
$\sigma _{21}^{\left( +\right) }$ and $\sigma _{21}^{\left( -\right)
}$ are the symmetric and antisymmetric superpositions of $\sigma
_{21}^{\mathrm{EIT}}$ and $\sigma _{21}^{\mathrm{Para}}$,
respectively.

Typically, if the amplitudes and phases of Rabi frequencies are
chosen as
\begin{equation}
\left\vert \Omega _{21}\right\vert =\left\vert \Omega
_{31}\right\vert,\Theta =0,  \label{Condition}
\end{equation}
one can check from Eq.~\eqref{sigma21} that $\mathrm{Im}\sigma
_{21}^{(-)}\simeq 0$ and $\mathrm{Im}\sigma _{21}^{(+)}$ has a peak
at the detuning $\Delta=-\left\vert \Omega _{32}\right\vert /2$;
while $\mathrm{Im}\sigma _{21}^{(+)}\simeq 0$ and $\mathrm{Im}\sigma
_{21}^{(-)}$ has a peak at the detuning $\Delta=\left\vert \Omega
_{32}\right\vert /2$ (see also Fig.~\ref{interference}). Namely, we
get chirality-dependent single-peak structures of the
absorption-related quantities $\mathrm{Im}\sigma _{21}^{(\pm)}$.
Therefore, by the locations of the absorption peaks one can mark the
left- and right handed molecules.

Physically, the above results can be further explained using the
dressed-state picture, see the inset in Fig.~\ref{interference}. For
left-handed molecules, under the strong driving the levels
$\left\vert 2\right\rangle _{\mathrm{L}}$ and $\left\vert
3\right\rangle _{\mathrm{L}}$\ are split into $\left\vert 2,\pm
\right\rangle _{\mathrm{L}}$ and $\left\vert 3,\pm \right\rangle
_{\mathrm{L}}$, respectively. The energy differences of each pair of
dressed sublevels is $\left\vert \Omega _{32}\right\vert $. When the
probe detuning is set as $\Delta =-\left\vert \Omega
_{32}\right\vert /2$, constructive interference between transition
channels $\left\vert 1\right\rangle _{\mathrm{L}}\leftrightarrow
\left\vert 2,-\right\rangle _{\mathrm{L}}$ and $\left\vert
1\right\rangle _{\mathrm{L}}\leftrightarrow \left\vert
3,-\right\rangle _{\mathrm{L}}\leftrightarrow \left\vert
2,-\right\rangle _{\mathrm{L}}$ can generate a characteristic peak
for left-handed molecules; while when $\Delta =\left\vert \Omega
_{32}\right\vert /2$, the transitions $\left\vert 1\right\rangle
_{\mathrm{L}}\leftrightarrow \left\vert 2,+\right\rangle
_{\mathrm{L}}$ and $\left\vert 1\right\rangle
_{\mathrm{L}}\leftrightarrow \left\vert 3,+\right\rangle
_{\mathrm{L}}\leftrightarrow \left\vert 2, +\right\rangle
_{\mathrm{L}}$ interfere destructively and thus $\mathrm{Im}\sigma
_{21}^{\left( +\right) }\simeq 0$. For the right-handed molecules
case, the explain can be given in the same way.

\begin{figure}[t]
\includegraphics[width=0.4\textwidth]{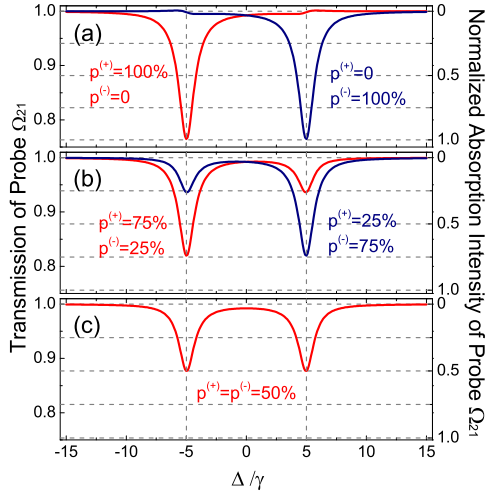}
\caption{(Color online) Transmission coefficient $T$ and normalized
absorption intensity $\tilde{I}=(1-T)/( h^{(+)}+h^{( -)})$ of the
probe field $\Omega_{21}$ as functions of detuning, with different
enantiomeric excess. The coherent driving fields enter the mediums
with $\Omega _{32}\left( 0\right) =10\protect\gamma $, $\Omega
_{21}\left( 0\right) =\Omega _{31}\left( 0\right) =\protect\gamma
/10$, satisfying the condition \eqref{Condition}. Optical depth is
set as $\protect\zeta =0.2$. The other parameters are the same as
those in Fig.~\ref{interference}. } \label{CP}
\end{figure}

Based on above qualitative analysis, we now specifically consider a
medium consisting of a mixture of left- and right-handed
($\Delta$-type) molecules. The optical waves propagating in the
medium is described by the Maxwell equation~\cite{Scully}:
$-\partial _{zz}E_{ij}+c^{-2}\partial _{tt}E_{ij}=-\mu _{0}\partial
_{tt}P_{ij}$. The induced oscillating polarizations $P_{ij}$ are
given by
$P_{ij}\left( z,t\right) =\frac{1}{2} \mathscr{P}_{ij}e^{-i\left(
\nu _{ij}t-k_{ij}z+\phi _{ij}\right) }+\mathrm{c.c.}$,
with slowly varying polarization
$\mathscr{P}_{ij}=2N\mu _{ji}( p^{(+) }\sigma
_{ij}^{(+)}+p^{(-)}\sigma _{ij}^{(-)})$,
where $N$ is the total molecular density, $p^{\left( \pm\right) }$
are the percentages of the two enantiomers.
In the slowly-varying amplitude and phase approximation, the Maxwell
equation are reduced to
\begin{equation}
\frac{\partial \Omega _{ij}}{\partial z}=i\frac{N\mu _{ji}^{2}\nu
_{ij}}{\hbar \epsilon _{0}c}\left( p^{(+) }\sigma
_{ij}^{(+)}+p^{(-)}\sigma _{ij}^{(-)}\right) e^{-i\theta _{ij}}.
\label{Maxwelleq}
\end{equation}
In addition, we can neglect the change of the strong control field
$\Omega _{32}$, and approximately express $\sigma _{31}^{(\pm)}$ and
$\sigma _{21}^{(\pm)}$ in the first order of $\Omega _{31}$ and
$\Omega _{21}$. Thus the probes $\Omega _{31}$ and $\Omega _{21}$
obey the following propagation equations
\begin{subequations}
\begin{eqnarray}
\frac{\partial \Omega _{31}}{\partial z} &=&i\frac{N\mu _{13}^{2}\nu _{31}}{%
Z\hbar \epsilon _{0}c}\left( -\frac{1}{4}\Omega _{21}\Omega _{32}\delta p+%
\frac{1}{2}i\lambda _{21}\Omega _{31}\right)  \label{approeq31}
\\
\frac{\partial \Omega _{21}}{\partial z} &=&i\frac{N\mu _{12}^{2}\nu _{21}}{%
Z\hbar \epsilon _{0}c}\left( -\frac{1}{4}\Omega _{31}\Omega _{32}^{\ast
}\delta p+\frac{1}{2}i\lambda _{31}\Omega _{21}\right)  \label{approeq21}
\end{eqnarray}
\end{subequations}
with $\delta p=p^{\left( +\right) }-p^{\left( -\right) }$ being the
enantiomeric difference. To achieve desirable chirality
identification, the three coherent fields should satisfy the
condition~\eqref{Condition} at $z=0$. Furthermore, we define the
transmission coefficient of the probe $\Omega _{21}$ as
$T=\left\vert \Omega _{21}\left( z \right) \right\vert
^{2}/\left\vert \Omega _{21}\left( 0\right) \right\vert ^{2}$ and
the corresponding absorption intensity $I=1-T$. Additionally, if
assuming $|\Omega _{32}|\gg\gamma_{12},  \gamma_{13}$, then at
$\Delta =\mp \left\vert \Omega _{32}\right\vert /2$, we have
$Z\simeq\pm i |\Omega _{32}|(\gamma_{12}+\gamma_{13})/2$,
$\lambda_{31}\simeq\lambda_{21}\simeq\pm i |\Omega_{32}|/2$.
Consequently, the heights of characteristic peaks $h^{\left( \pm
\right) }$ (i.e., the absorption intensity at $\Delta =\mp
\left\vert \Omega _{32}\right\vert /2$) in the absorption spectrum
can be gotten by solving Eqs.~\eqref{approeq31}-~\eqref{approeq21}:
\begin{eqnarray}
&&h^{\left( \pm \right) }=1-\frac{1}{4B}e^{-C\zeta }  \notag \\
&&\times \left[ \left( 1-A\pm 2\delta p\right) \left( 1-e^{D\zeta }\right) +%
\sqrt{B}\left( 1+e^{D\zeta }\right) \right] ^{2}  \label{Cpeak}
\end{eqnarray}
with
$\zeta =N\mu _{21}^{2}\nu _{21}z/(\hbar \epsilon _{0}c\Gamma
_{21})$,
$A=\mu _{31}^{2}\nu _{31}/(\mu _{21}^{2}\nu _{21})$,
$B=( 1-A) ^{2}+4A(\delta p) ^{2}$,
$C=\Gamma _{21}( 1+A+\sqrt{B})/[2( \gamma _{12}+\gamma _{13}]$,
$D=\Gamma _{21}\sqrt{B}/[2\left( \gamma _{12}+\gamma _{13}\right)]$.
Typically, in the optical thin region (i.e., the dimensionless
optical depth $\zeta\ll 1$) the height of each characteristic peak
can be approximately expressed as
\begin{equation}
h^{\left( \pm \right) }\simeq \frac{2\Gamma _{21}\zeta }{\gamma _{12}+\gamma
_{13}}p^{\left( \pm \right) },  \label{linearrelation}
\end{equation}
which is linearly proportional to the percentage of according chiral
molecules.
Therefore, if we define the normalized heights of characteristic
peaks as $\tilde{h}^{\left( \pm \right) }=h^{\left( \pm \right)
}/\left( h^{\left( +\right) }+h^{\left( -\right) }\right) $, then
$\tilde{h}^{\left( \pm \right) }\simeq p^{\left( \pm \right) }$ in
the linear range \ $\zeta \ll 1$.

\begin{table}[b]
\caption{Relation between $\protect\delta p^{\prime }$ and
$\protect\delta p$ in the linear range of optical depth
$\protect\zeta $, for certain typical parameters of $\protect\zeta $
and $\left\vert \Omega _{32}\right\vert $. The other parameters are
the same as those in Fig.~\ref{CP}.}
 \label{linear}
\begin{ruledtabular}
\begin{tabular}{cccccccc}
&&$\zeta=0.05$&$\zeta=0.1$&$\zeta=0.2$\\
&$\delta p$ (\%)&$\delta p'$ (\%)&$\delta p'$ (\%)&$\delta p'$ (\%)&$\left\vert \Omega _{32}\right\vert/\gamma$\\
\hline
&0   &0&0&0\\
&$\pm25$&$\pm24.44$&$\pm24.21$&$\pm23.75$\\
&$\pm50$&$\pm48.96$&$\pm48.59$&$\pm47.85$&10\\
&$\pm75$&$\pm73.66$&$\pm73.33$&$\pm72.66$\\
&$\pm100$&$\pm98.64$&$\pm98.62$&$\pm98.59$\\
\\
&0   &0&0&0\\
&$\pm25$&$\pm24.77$&$\pm24.53$&$\pm24.07$\\
&$\pm50$&$\pm49.62$&$\pm49.25$&$\pm48.50$&100\\
&$\pm75$&$\pm74.66$&$\pm74.34$&$\pm73.67$\\
&$\pm100$&$\pm99.99$&$\pm99.99$&$\pm99.99$\\
\end{tabular}
\end{ruledtabular}
\end{table}

The above analysis can be verified by numerical solving the master
equations~\eqref{Mastereq1}-\eqref{Mastereq5} and the relevant
Maxwell equation~\eqref{Maxwelleq}. Fig.~\ref{CP} presents the probe
transmission $T$ and the normalized absorption intensity $\tilde{I}
=I/\left( h^{\left( +\right) }+h^{\left( -\right) }\right) $ at
optical depth $\zeta =0.2$ versus the detuning $\Delta$. It is seen
that: (1) At detuning $\Delta =\mp \left\vert \Omega
_{32}\right\vert /2$, there exist two characteristic peaks
corresponding to left- and right-handed molecules, respectively; (2)
The percentages of the enantiomers are directly related to the
heights of the characteristic peaks. Furthermore, Table \ref{linear}
shows also that, in the linear region the enantiomeric difference
$\delta p$ can be read directly from the difference of the heights
of two characteristic peaks $\delta p^{\prime }=\tilde{h}^{\left(
+\right) }-\tilde{h}^{\left( -\right) }$, namely, $\delta p\simeq
\delta p^{\prime }$. Table \ref{linear} also tell us that for
stronger $\Omega _{32}$ and smaller $\zeta $, $\delta p^{\prime }$
can more accurately reflect the values of enantiomeric difference.

\begin{figure}[t]
\includegraphics[width=0.5\textwidth]{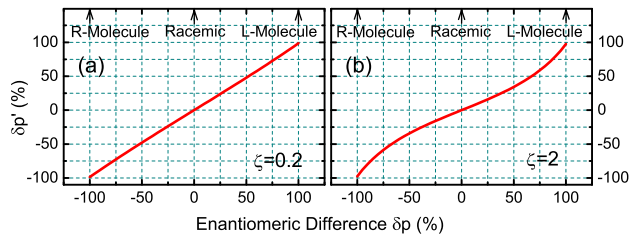}
\caption{(Color online) The difference of the heights of two
characteristic
peaks $\protect\delta p^{\prime }=\tilde{h}^{\left( +\right) }-\tilde{h}%
^{\left( -\right) }$ as a fuction of enantiomeric difference $\protect\delta %
p=p^{\left( +\right) }-p^{\left( -\right) }$ at differnt optical depth: (a) $%
\protect\zeta =0.2$; (b) $\protect\zeta =2$. The other parameters
are the same as those in Fig.~\protect\ref{CP}. } \label{readpeak}
\end{figure}

In Fig.~\ref{readpeak}(a), we show that, for relatively-weak optical
depths, e.g., $\zeta =0.2$, an approximately linear relation exists
between the characteristic peak difference $\delta p^{\prime }$ and
the enantiomeric difference $\delta p$. When $\zeta$ increases,
$\delta p^{\prime }$ is never linearly proportional to $\delta p$,
see specifically Fig.~\ref{readpeak}(b). For this case we can not
read directly the percentages of the chiral molecules from the
heights of characteristic peaks. However, note that whether in
linear regime or not, $\delta p^{\prime }$ is always a monotonic
function of $\delta p$. Thus, for a given $\delta p^{\prime }$
(attained by measuring the absorption peaks) the percentage of two
enantiomers can always be accurately identified from the $\delta
p^{\prime }$-$\delta p$ curve.

\textit{Discussions and conclusions---}Coherence effects, such as
EIT, coherent population trapping, lasing without inversion, and so
on, was firstly observed in atomic systems~\cite{CPT,EIT,LWI}, and
recently have been demonstrated in certain molecular
systems~\cite{EITinMolecule1,EITinMolecule2,EITinMolecule3}.
Typically, to realize these effects in molecular system should
overcome the relatively weak optical response of the molecules. This
can be achieved in, such as the hollow-core photonic crystal fiber
(HC-PCF). In this system atoms and molecules can be confined in the
core and thus the light-matter interactions can be enhanced
significantly~\cite{HCPCF}. Indeed, HC-PCF filled with molecular gas
have been demonstrated for significantly enhancing various coherence
effects, such as EIT~\cite{EITinMolecule2,EITinMolecule3} and
four-wave mixing~\cite{FWMinPCF}.
On the other hand, laser cooled ensemble containing few atoms
trapped inside the HC-PCF has been achieved to demonstrate the EIT
related effects~\cite{ColdAtomInPCF}. Hopefully, similar experiments
to confine few cold molecules could also be implemented to realize
the desirable coherence effects. Thus, if our coherence effect
relevant protocol is implemented based on cold molecular ensemble in
HC-PCF, an effective method to probe chirality of a small amount of
molecular sample is possible.

Although our calculations and discussions are focused on the cold
molecules case, our protocol can still be applied to probe the
chirality of hot molecules. Certainly, including the Doppler shifts
due to the thermal motion of molecules, $\lambda _{ij}$ should be
redefined as $\lambda _{ij}=\gamma _{ij}+i\left( k_{ij}v_{z}+\Delta
_{ij}\right)$. Here, $v_{z}$ is the $z$ component of the
translational velocity of the molecules. The according medium
polarization components is then obtained by integrating the density
matrix elements over the molecular distribution on velocities, i.e.,
$\tilde{\sigma}_{ij}=\left( \pi u_{D}^{2}\right)
^{-1/2}\int_{-\infty }^{+\infty }\sigma _{ij}\left( v_{z}\right)
\exp \left( -v_{z}^{2}/u_{D}^{2}\right) dv_{z}$ with $u_{D}$  the
most probable velocity of molecules.
This implies that the Doppler shifts will lead to the broadening of
spectral lines and may obscure the characteristic peaks. However, a
stronger control field $\Omega _{32}$ could be applied to assure the
Rabi split is larger than the Doppler-broadening, and consequently
similar chirality-related spectra could be observed.

In summary, we have proposed an quantum optical approach to probe
molecular chirality based on phase-dependent quantum coherence and
interference effects. The molecular chirality could be identified by
the chirality-dependent information in the absorption spectrum. Our
method may provide a sensitive chirality-probe and lead to
applications in organic chemistry, biochemistry, and pharmacology.

This work was supported in part by National Natural Science Foundation of
China under Grant Nos. 10874142, 90921010, and the National Fundamental
Research Program of China through Grant No. 2010CB923104.





\end{document}